\documentstyle[epsfig]{mn}

\newif\ifAMStwofonts
\ifoldfss
 \ifCUPmtlplainloaded \else
 \NewTextAlphabet{textbfit} {cmbxti10} {}
 \NewTextAlphabet{textbfss} {cmssbx10} {}
 \NewMathAlphabet{mathbfit} {cmbxti10} {} % for math mode
 \NewMathAlphabet{mathbfss} {cmssbx10} {} % " " "
 \fi
 \ifAMStwofonts
 \ifCUPmtlplainloaded \else
 \NewSymbolFont{upmath} {eurm10}
 \NewSymbolFont{AMSa} {msam10}
 \NewMathSymbol{\upi} {0}{upmath}{19}
 \NewMathSymbol{\umu} {0}{upmath}{16}
 \NewMathSymbol{\upartial}{0}{upmath}{40}
 \NewMathSymbol{\leqslant}{3}{AMSa}{36}
 \NewMathSymbol{\geqslant}{3}{AMSa}{3E}

 \let\leq=\leqslant \let\le=\leqslant
  
 \fi
 \fi
\fi % End of OFSS

\ifnfssone
 \newmathalphabet{\mathit}
 \addtoversion{normal}{\mathit}{cmr}{m}{it}
 \addtoversion{bold}{\mathit}{cmr}{bx}{it}
 \newmathalphabet{\mathbfit} % math mode version of \textbfit{..}
 \addtoversion{normal}{\mathbfit}{cmr}{bx}{it}
 \addtoversion{bold}{\mathbfit}{cmr}{bx}{it}
 \newmathalphabet{\mathbfss} % math mode version of \textbfss{..}
 \addtoversion{normal}{\mathbfss}{cmss}{bx}{n}
 \addtoversion{bold}{\mathbfss}{cmss}{bx}{n}
 \ifAMStwofonts
 \ifCUPmtlplainloaded \else
 \UseAMStwoboldmath
 \makeatletter
 \new@mathgroup\upmath@group
 \define@mathgroup\mv@normal\upmath@group{eur}{m}{n}
 \define@mathgroup\mv@bold\upmath@group{eur}{b}{n}
 \edef\UPM{\hexnumber\upmath@group}
 \new@mathgroup\amsa@group
 \define@mathgroup\mv@normal\amsa@group{msa}{m}{n}
 \define@mathgroup\mv@bold\amsa@group{msa}{m}{n}
 \edef\AMSa{\hexnumber\amsa@group}
 \makeatother
 \mathchardef\upi="0\UPM19
 \mathchardef\umu="0\UPM16
 \mathchardef\upartial="0\UPM40
 \mathchardef\leqslant="3\AMSa36
 \mathchardef\geqslant="3\AMSa3E

 \let\leq=\leqslant \let\le=\leqslant

 \fi
 \fi
\fi % End of NFSS release 1

\ifnfsstwo
 \DeclareMathAlphabet{\mathbfit}{OT1}{cmr}{bx}{it}
 \SetMathAlphabet\mathbfit{bold}{OT1}{cmr}{bx}{it}
 \DeclareMathAlphabet{\mathbfss}{OT1}{cmss}{bx}{n}
 \SetMathAlphabet\mathbfss{bold}{OT1}{cmss}{bx}{n}
 \ifAMStwofonts
 \ifCUPmtlplainloaded \else
 \DeclareSymbolFont{UPM}{U}{eur}{m}{n}
 \SetSymbolFont{UPM}{bold}{U}{eur}{b}{n}
 \DeclareSymbolFont{AMSa}{U}{msa}{m}{n}
 \DeclareMathSymbol{\upi}{0}{UPM}{"19}
 \DeclareMathSymbol{\umu}{0}{UPM}{"16}
 \DeclareMathSymbol{\upartial}{0}{UPM}{"40}
 \DeclareMathSymbol{\leqslant}{3}{AMSa}{"36}
 \DeclareMathSymbol{\geqslant}{3}{AMSa}{"3E}

 \let\leq=\leqslant \let\le=\leqslant

 \fi
 \fi
\fi % End of NFSS release 2

\ifCUPmtlplainloaded \else
 \ifAMStwofonts \else % If no AMS fonts
 \def\upi{\pi}
 \def\umu{\mu}
 \def\upartial{\partial}
 \fi
\fi

\title[Distant field BHB stars and the mass of the Galaxy I]
{Distant field BHB stars and the mass of the Galaxy I:
Classification of halo A--type stars \thanks{Based on observations
obtained at the Jacobus Kapteyn Telescope, the Isaac Newton Telescope,
and the William Herschel Telescope, La Palma, and the Anglo--Australian
Telescope, Siding Spring Observatory, Australia.}}

\author[L. Clewley, et al.]
{L. Clewley,$^{1}$ S. J. Warren,$^{1}$ P. C. Hewett,$^{2}$
John E. Norris$^{3}$, R. C. Peterson$^{4}$,
\newauthor N. W. Evans$^{5}$\\
$^{1}$Blackett Laboratory, Imperial College of Science Technology
and Medicine, Prince Consort Rd, London SW7 2BW \\
$^{2}$Institute of Astronomy, Madingley Road, Cambridge CB3 0HA\\
$^{3}$Research School of Astronomy \& Astrophysics, The Australian National
University,\\ Mount Stromlo Observatory, Cotter Road, Weston, ACT 2611,
Australia\\
$^{4}$UCO/Lick Observatory, Department of Astronomy, University of
California, Santa Cruz, Santa Cruz, CA 95064, USA\\
$^{5}$Theoretical Physics, 1 Keble Road, Oxford OX1 3RH}

\date{Accepted
 Received
 in original form}

\begin{document} \maketitle \begin{abstract} This is the first in a
series of three papers presenting a new calculation of the mass of the
Galaxy based on radial velocities and distances measured for a sample
of some 100 faint $16<B<20$ field blue horizontal-branch (BHB) stars.
This study aims to reduce the uncertainty in the measured mass of the
Galaxy by increasing the number of halo objects at Galactocentric
distances $r>30$ kpc with measured radial velocities by a factor five.
Faint A-type stars in the Galactic halo have been identified from
$UB_JR$ photometry in six UK Schmidt fields. These samples include
field BHB stars as well as less luminous stars of main-sequence
surface gravity, which are predominantly field blue stragglers. We
obtain accurate CCD photometry and spectra to classify these
stars. This paper describes our methods for separating out clean
samples of BHB stars in a way that is efficient in terms of telescope
time required. We use the high signal--to--noise ratio (S/N) spectra
of A--type stars of Kinman, Suntzeff \& Kraft (published in 1994), and
their definitive spectrophotometric $\Lambda$ classifications, to
assess the reliability of two methods, and to quantify the S/N
requirements. First we revisit, refine and extend the hydrogen line
width {\em versus} colour relation as a classifier (here called the
{\em $D_{0.15}$--Colour} method). The second method is new and
compares the {\em shapes} of the Balmer lines.  With this method (here
called the {\em Scale width--Shape} method) there is no need for
colours or spectrophotometry. Using the equivalent width of the Ca II
K line as an additional filter we find we can reproduce Kinman,
Suntzeff \& Kraft's $\Lambda$ classifications with both methods.  In a
sample of stars with strong Balmer lines, EW H$\gamma>13$\AA\,
(equivalent to the colour range $0 \leq (B-V)_0 \leq 0.2$), halo BHB
stars can be separated from halo blue stragglers reliably. For the
spectroscopy (i.e. both classification methods) the minimum required
continuum S/N is $15\,{\mathrm\AA}^{-1}$. For the {\em
$D_{0.15}$--Colour} method $(B-V)_0$ colours accurate to 0.03 mag. are
needed.  \end{abstract}

\section{Introduction}
The mass of the Galaxy is a key quantity for our understanding of the
nature and distribution of dark matter. Although we know a great deal
about the contents and properties of the halo of the Galaxy our
current estimate of the total mass is subject to a large
uncertainty. In part this is due to the Sun's location within the disk
of the Galaxy which means it is difficult to determine the rotation
curve accurately outside the solar radius. Beyond $\sim 20\,$kpc
measures of the enclosed mass of the halo have relied principally on
the kinematics of satellite galaxies and globular clusters. The
current state of the art is the analysis by Wilkinson \& Evans (1999,
hereafter, WE99) who calculate the mass within $50\,$kpc to be
$5.4^{+0.2}_{-3.6} \times 10^{11} M_{\odot}$, and the total mass to be
$1.9^{+3.6}_{-1.7} \times 10^{12} M_{\odot}$. The quoted uncertainties
are larger than in earlier studies (e.g. Little \& Tremaine, 1987,
Zaritsky et al. 1989, Kochanek, 1996) and were determined by Monte
Carlo simulation of artificial data sets. The two principal sources of
error are i) the large uncertainties in the proper motions for the six
satellites for which measures exist, and ii) the small size of the
dataset \---\ there are currently only 27 known satellites at
Galactocentric distances $> 20\,$kpc.  Therefore, we are in the
peculiar situation where the mass profile of the Galaxy is less well
determined than for some nearby spiral galaxies. There is no
possibility of substantially increasing the number of known satellite
galaxies and globular clusters and the best prospect for improving the
measurement of the mass is through isolating large numbers of another
distant halo--tracer. Field blue horizontal branch (BHB) stars should
provide just such a tracer and WE99 calculate that to reduce the
uncertainty on the total mass to $20\%$ requires a sample of 200
distant BHB stars.

This is the first in a series of three papers presenting a new
calculation of the mass of the Galaxy using radial velocities of BHB
stars. Field BHB stars are luminous standard candles that are abundant
in the Galactic halo (e.g. Yanny et al. 2000), and for nearly twenty
years, since the study of Pier (1983), have presented an
under--exploited resource with which to measure the density profile
and phase space structure of the Galaxy halo out to large distances,
$\sim 100\,$kpc. A number of dynamical analyses of rather small
samples of BHB stars have been published (e.g. Sommer--Larsen,
Christensen \& Carter, 1989, Norris \& Hawkins, 1991, Arnold \&
Gilmore, 1992). Unfortunately, samples of field A--type stars in the
halo include not only BHB stars but also stars of main sequence
surface gravity, field blue stragglers, that are some 2 magnitudes
less luminous. Progress towards the goal of acquiring a large sample
of distant BHB stars has been slow because of the difficulty of
separating out the BHB stars without the investment of large amounts
of telescope time. In this first paper we describe our procedures for
classifying samples of halo A--type stars, and present a new efficient
method that requires only spectroscopic observations of intermediate
signal--to--noise ratio. In Paper II we will present photometry and
spectroscopy of faint $16<B<20$ candidate BHB stars in two northern
high Galactic latitude fields and four southern fields. Paper III will
contain the dynamical analysis of the new sample of confirmed BHB
stars and a new estimate of the mass of the Galaxy.

The structure of this paper is as follows. In \S2 we provide a brief
outline of the basic parameters of our new survey for distant field BHB
stars in the halo. \S3 contains a summary of previous methods employed
to sieve out BHB stars from samples of halo A--type stars. In \S4 we
present the details of the two classification methods we employ, called
the {\em $D_{0.15}$--Colour} method, and the {\em Scale width--Shape}
method. We describe the procedures used to measure the Balmer line
profiles and the equivalent width of the Ca II K line, and quantify the
random and systematic errors. In \S5 we use the high signal--to--noise
ratio (S/N) spectra of Kinman, Suntzeff \& Kraft (1994; hereafter KSK)
and their spectrophotometric classifications to quantify the S/N
requirements for applying these methods to samples of faint halo stars.
In \S6 we discuss the use of the Ca II K line as a metallicity
indicator. Finally, \S7 provides a summary of the main conclusions of
the paper.

\section{The survey} In a $UBV$ two--colour plot (or its equivalent
such as $ugr$ or $UB_JR$) of a high Galactic latitude field, halo
A--type stars are identifiable as a faint continuation of the spectral
sequence to hotter types, beyond the main sequence turnoff at spectral
type F, with colours $0.0<(B-V)_0 <0.2$ (see e.g. Yanny et al. 2000;
Fig. 1). As illustrated in striking fashion by Yanny et al. using
Sloan Digital Sky Survey (SDSS) data, these A--type stars include not
only luminous field BHB stars, of absolute magnitude $M_V\sim 0.7$,
but also stars of main--sequence surface gravity that are some two
magnitudes less luminous. The nature of these stars is not entirely
clear but recent work supports the notion that the majority are blue
stragglers in binary systems created by mass transfer when the
companion star overfills its Roche lobe (Preston \& Sneden 2000). As
noted by Preston, Beers \& Shectman (1994) this means that in the
field blue stragglers are much more common than BHB
stars, whereas in globular clusters they are usually rarer. This
difference could be explained by the destruction of wide binaries in
globular clusters. This picture is supported by Carrera et al. (2002)
who find an intermediate value of the ratio of the numbers of
main--sequence--gravity A stars and BHB stars for the
Ursa Minor dwarf galaxy, which has a stellar density intermediate
between that of the field and that of a globular cluster. For these
reasons in the remainder of this paper we refer to these distant halo
A--type stars with main--sequence surface gravities as field blue
stragglers.

The combination of their substantial luminosity, ensuring they can be
detected to large distances, and their very small spread in absolute
magnitude, making them effective standard candles, means BHB stars are
ideal dynamical tracers. For the same apparent magnitude the field
blue stragglers are a factor 2.5 less distant and it is therefore
crucial to separate the two populations. Our survey employs APM scans
of UK Schmidt Telescope photographic plates. We use pairs of plates in
each of the $U$, $B_J$, and $R$ bands, in six fields. The coordinates
of the six survey fields are provided in Table 1. The fields were
selected to give good coverage in opposing directions above and below
the Galactic plane, with a range of Galactic longitudes, subject to
the availability of suitable plate material in the UK Schmidt
Telescope archive \---\ which effectively restricts the search to
negative declinations.  There was a preference for high Galactic
latitudes to minimize extinction. The different lines of sight through
the Galaxy provide complementary information on the anisotropy of the
stellar orbits. Further details of the plate material, observing
programme and data processing are provided in Paper II. With plate
material from the UK Schmidt Telescope A--type stars can be reliably
separated from F stars down to $B \simeq 20$. We obtain accurate CCD
BV photometry and spectra at $\sim 3$\AA\,FWHM resolution to classify these
stars. As described in Paper II we select candidate BHB stars by
colour in the magnitude range $16<B<20$, which corresponds (assuming
$M_V\sim 0.7$) to the distance range $11-72\,$kpc for BHB stars, but
to only $5-29\,$kpc for the field blue stragglers. Pulsating RR Lyrae
stars have also provided a
fruitful source of HB standard candles (Hawkins 1984, Ivezi\'{c} et al. 2000). These stars reside in the instability strip,
redward of B-V = 0.2, with $\Delta$B $\approx$ 1 mag. and with periods
around 0.5 days. However, a number of repeat observations are required to isolate these stars, making it time consuming to gather
observational data. Red horizontal branch stars, which are also
standard candles, have colours $(B-V)_0>0.4$, similar to the colours
of F stars, and therefore cannot be identified effectively using
broadband photometry.

\begin{table}
 \centering
 \begin{tabular}{@{}lrrrr}
 \hline \\[-12pt]
	Field name & \multicolumn{1}{c}{l} & \multicolumn{1}{c}{b} &
 \multicolumn{1}{c}{RA} & \multicolumn{1}{c}{Dec} \\
	 & & &
 \multicolumn{2}{c}{J 2000} \\
	\hline \\[-12pt]
SGP & 250 & -89 & 0 55 & -27 47 \\
SA94 & 175 & -50 & 2 53 & 0 12 \\
F358 & 236 & -54 & 3 38 & -34 50 \\
F854 & 244 & 45 & 10 23 & -0 15 \\
F789 & 299 & 58 & 12 43 & -5 16 \\
MT & 37 & -51 & 22 06 & -18 39 \\
 \hline
\end{tabular}
\label{fields}
\caption{Galactic and equatorial coordinates of the six survey
fields, ordered by Right Ascension.} 
\end{table}

\section{Previous BHB star classifications}

To classify A stars the problem is to separate out the giants,
i.e. the BHB stars, from the dwarfs, i.e. main sequence A stars or
blue stragglers, using an indicator that depends on surface
gravity. In samples of relatively nearby stars the dwarfs (i.e. main
sequence A stars in the disk in this case) greatly outnumber the BHB
stars, and also the two populations have very different kinematic
properties. In such circumstances, especially for a dynamical study,
an extremely reliable technique is required to ensure contamination of
the BHB--sample is minimized, and high-resolution spectroscopy is
necessary (Kinman et al. 2000). Fortunately, in the outer halo, far
from the disk, the number of BHB stars and potential contaminants
(here blue stragglers) are comparable. Furthermore the two populations
will have much more similar kinematic properties, so the problem is
somewhat less demanding. Our goal is to identify an efficient method
for the classification of BHB stars in the halo that is largely
complete while also ensuring the contamination from non--BHB stars is
$\la 10\%$. Under these requirements high resolution spectroscopy is
not necessary.

The study by KSK of the problem of classifying distant A--type stars
in the Galactic halo has provided the benchmark for future work.  KSK
studied three surface gravity indicators, each as a function of
$(B-V)_0$ colour: i) the strength of the Balmer jump, which may be
quantified by the Stromgren index $c_1$, or some similar close
equivalent, ii) the steepness of the Balmer jump, quantified by the
spectrophotometric $\Lambda$ parameter, iii) the width of the Balmer
lines. (Note that broadband $UBV$--photometry on its own provides only
a crude separation of the populations and was not considered.) Of the
three methods the $\Lambda$ parameter provides the cleanest separation
and is the preferred indicator. However, measurement of the $\Lambda$
parameter is impractical for faint stars. A S/N$\sim
40\,{\mathrm\AA}^{-1}$\footnote{In this paper all values of S/N are
quoted as ${\mathrm\AA}^{-1}$ by dividing the S/N per pixel by the
square root of the pixel size in ${\mathrm\AA}$.} is required, which,
at $V=19$, takes some two hours of integration on a 4m telescope in
dark time.  Furthermore, a wide slit is required for
spectrophotometry, necessitating the acquisition of a second spectrum,
with a narrower slit, to measure the star's radial velocity
accurately.  Method i) is also impractical for very similar reasons.

The third method, based on the width of the Balmer lines as a function
of $(B-V)_0$ colour, is the most efficient, since the spectrum also
provides the radial velocity. This method has been used in all
previous studies of the dynamics of faint BHB stars. Flynn,
Sommer--Larsen \& Christensen (1994) classified a sample of A--type
stars using the Stromgren index $c_1$ and demonstrated that the stars
classified as BHB also cleanly separate out in a plot of line--width
against $(B-V)_0$ colour. Similarly, KSK found that stars classified
BHB by the $\Lambda$ parameter separate from blue stragglers on the
basis of line--width.  However, neither Flynn et al. nor KSK
quantified the spectroscopic S/N required or the fractional
contamination of BHB samples at a given S/N.  Finally, Wilhelm, Beers
\& Gray (1999a) made a detailed study of the problem, adding $U$
photometry to the $BV$ photometry and line--widths. Nevertheless,
although the $U-B$ colour is a useful indicator of surface gravity for
redder colours $0.2<(B-V)_0<0.4$, it adds less information over the
range $0.0<(B-V)_0<0.2$.

\section {New techniques for classifying faint halo A-type stars}

In this section we describe new techniques for classifying faint halo
A-type stars efficiently. In considering the best way to separate BHB
stars and blue stragglers we first revisited the use of the line
width {\em versus} colour relation as a discriminant. We have made a
number of small refinements over the work of KSK: i) we use a
functional fit to measure the line widths, removing subjectivity and
allowing us to quantify the uncertainties, ii) we use a slightly
different measure of line width than KSK, called $D_{0.15}$, iii) we
use the improved reddening corrections of Schlegel, Finkbeiner \&
Davis (1998), and iv) we use the Ca II K absorption line EW as an
additional classification filter. In the rest of this paper we refer
to this method as the {\em $D_{0.15}$--Colour} method.

In the course of this work we discovered that the dependence of the
profile (i.e. the detailed shape) of the Balmer lines on both
temperature and surface gravity is measurable. This led to what we
have called the {\em Scale width--Shape} classification method. {\em
With this method there is no need to obtain colours or spectrophotometry.}
These two classification methods are the subject of the remainder of
the paper.  Note that the two methods are not entirely independent
since they both use parameters derived from a fit to the line
shape. In \S5 we consider the circumstances under which one or other
is more useful.
 
In this section we summarise details of the KSK--sample of spectra
used in developing the classification methods. We then
describe the procedure for measuring the profiles of the Balmer lines,
and its application to the two methods of classification. Lastly we
explain the procedure for measuring the Ca II K line EW, and the uses
of this measure.

\subsection{The KSK data}
The high S/N spectra of stars analysed by KSK were kindly made
available in electronic form by Dr Nick Suntzeff. The total sample
includes 214 stars ranging in spectral type from B to F. The majority
of the sample are A--type stars, including BHB stars, main--sequence A
stars, and blue stragglers.

The  methods   of  classification  we  apply  are   effective  in  the
temperature range  where the  Balmer lines are  strong and do not work
blueward of $(B-V)_0=0$. Because the
{\em Scale width--Shape} method does not use colours, but both methods
use  spectroscopy, we need  a spectroscopic  criterion, rather  than a
colour  criterion,  for  defining  the  type  of  star  to  which  the
classification methods  apply. We select  all stars with  EW H$\gamma>
13\,$\AA. This sample contains 131 stars. We refer to this as the `KSK
total sample'. The EW limit corresponds approximately  to the colour
range $0<(B-V)_0<0.2$. The EW limit was chosen by trial and error to
produce the largest clean samples of BHB stars. With a lower limiting
EW we include stars with $(B-V)_0<0$ and $(B-V)_0>0.2$ where
discrimination is more difficult. With a higher limiting EW we remove
BHB stars from the sample but retain most of the blue stragglers.
 
The  main-sequence  stars  in  this  sample  are  typically  of  solar
metallicity. Since we are interested in classifying lower--metallicity
halo  stars  we  focus  in   particular  on  a  subsample  of  fainter
high--Galactic latitude  stars, consisting of 66  stars, of magnitudes
$13.0 \leq V \leq 16.5$ in two fields, SA57 at the Northern Polar Cap,
and RR7 in the direction of the Galactic anticentre. These stars have
 S/N in the range 23 to 80 ${\mathrm\AA}^{-1}$, with a mean of 66 ${\mathrm\AA}^{-1}$. We
refer  to this  subsample as  the `KSK  halo sample'.   The  KSK total
sample is  useful for showing  the broad trends in  the classification
parameters, because  with the larger  sample the sequences of  the two
populations are  better defined (e.g.   Fig.  \ref{colour_width}), but
because of the issue of metallicity we use the KSK halo sample only in
defining  the  classification  boundaries  in plots  of  the  relevant
parameters.

In comparing line widths measured for the same star in the KSK spectra
observed in different runs, we discovered a small systematic difference
between the spectra in the two halves of the dataset. The origin of
the difference was traced, with the assistance of Dr Suntzeff, to the
details of the flux--calibration procedures employed for different
observing runs.  We were able to correct the half of the data in error
by comparing the KSK spectra of the flux standard, flux calibrated
by itself, to the original Massey et al. (1988) spectrum.

\subsection{Functional fit to the Balmer lines}\label{functional_fit}
The results of a functional fit to the Balmer lines are used in both
classification methods. The width $D_{0.15}$ is determined from the
fit, while the parameters of the fit themselves are used in the {\em
Scale width--Shape} method.

\subsubsection{Continuum fit}
Before fitting the Balmer lines each spectrum is normalised to the
continuum by fitting a polynomial of degree three
\footnote{in IRAF parlance this is a polynomial of order four} to
regions of continuum well away from the wings of the lines.  Because
the Balmer lines are so broad care must be taken in fitting the
continuum, and we spent some time experimenting with the degree of the
polynomial and with different wavelength ranges for fitting. The
intervals chosen for the fit were 3863--3868, 3902--3925, 4020--4048,
4146--4275, 4388--4494 {\AA}. While the final choice of procedure is
inevitably subjective this is not a concern provided other observers
who follow the same procedure obtain the same results, within the
errors. This will be the case provided there is no systematic trend of
the measured parameters with S/N.

To test for systematics we created artificial spectra by adding noise
to high quality spectra of four stars. For each star we created 1000
spectra of a specified S/N, for several values of S/N in the range 7 to
30.
At a given S/N we measured the lines in each artificial spectrum, and
then calculated the mean and the error on the mean for each measured
parameter. For no parameter, at any S/N, was the mean value of the
parameter inconsistent with the value measured for the original high
S/N spectrum. In other words we found no significant systematic 
errors associated with the continuum fit.

\begin{figure}
\rotatebox{0}{
\centering{
\scalebox{0.40}{
\includegraphics*[20,140][700,700]{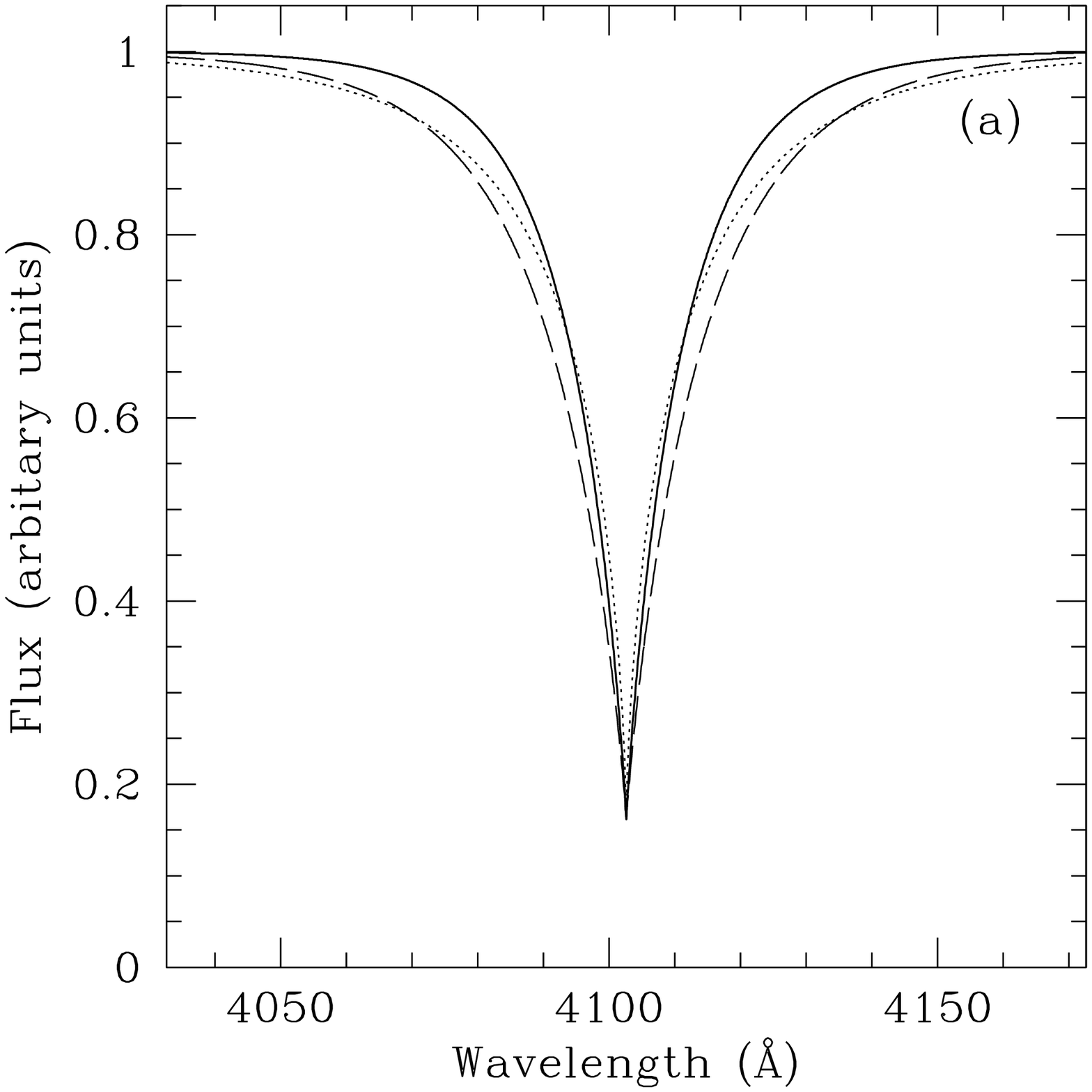}
}}}
%\end{figure}
%
%\begin{figure}
\rotatebox{0}{
\centering{
\scalebox{0.40}{
\includegraphics*[20,140][700,700]{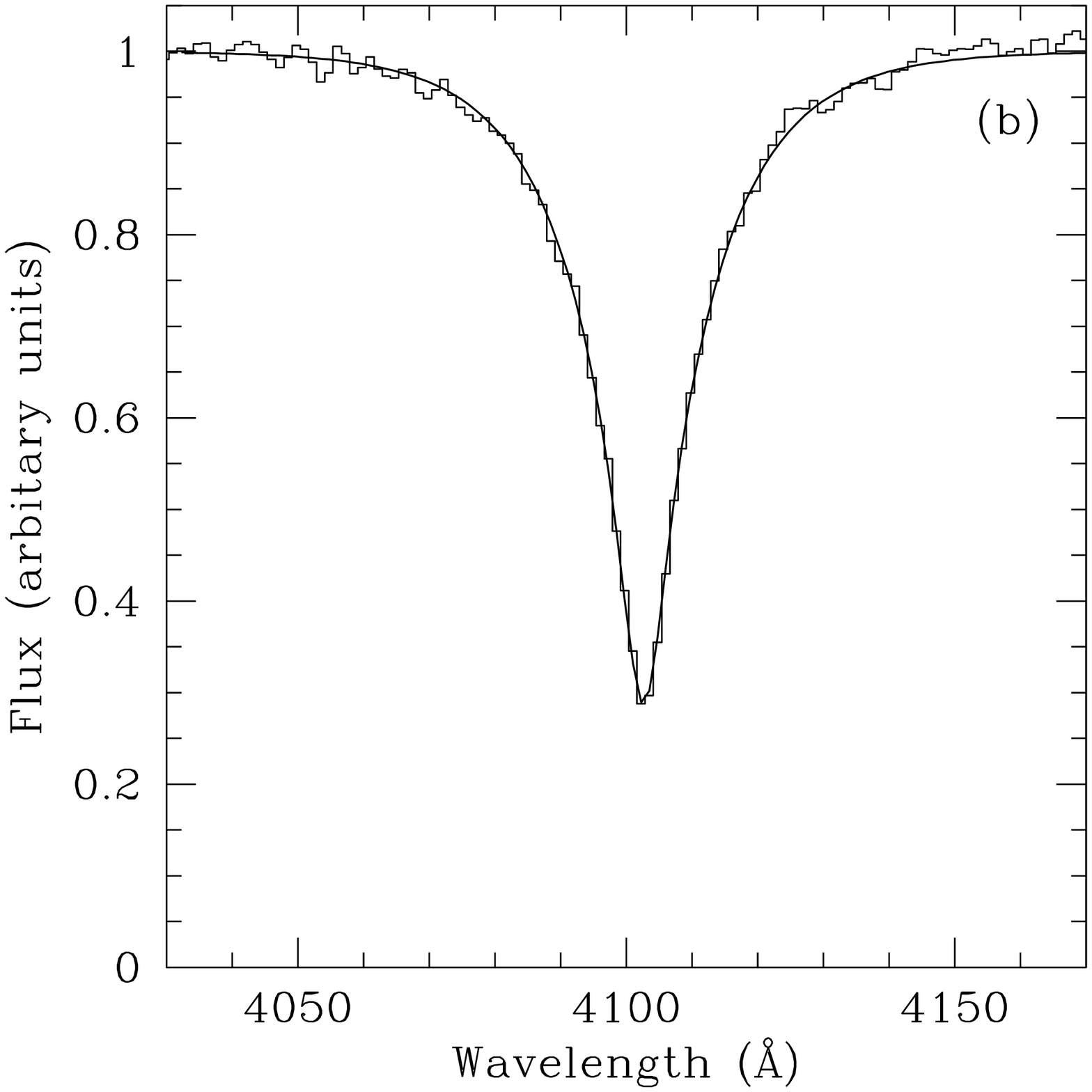}
}}}
\caption{(a) Illustration of the difference in line profile between
BHB stars and blue stragglers. The solid line plots the intrinsic
(i.e. deconvolved) profile of the star in (b) with $b=8.9$\AA\, and
$c=0.91$. The dashed line shows a blue straggler with the same
temperature $c=0.91$ and a broader scale width,
$b=12$\AA. Alternatively, the dotted line shows a blue straggler which
has the same scale width as the BHB star, $b=8.9$\AA, but is cooler,
$c=0.7$, and has broader line wings. (b) Convolved fit of the
H$\delta$ Balmer line for a typical BHB star in the KSK dataset. This
star has measured (deconvolved) parameters $b=8.9$\AA, $c=0.91$.}
\label{plot_fits}
\end{figure}

\subsubsection{Sersic profile fit}
To fit the line profiles in the normalized spectra we experimented
with several functions including an exponential, a $-5/2$ power law
(Pier, 1983), a general power--law, a Lorentzian, a Gaussian (Beers,
et al. 1992), and a Voigt profile (Wilhelm et al. 1999b). Finally we
adopted
the Sersic profile (Sersic, 1968), which is the exponential of a power
law:

\[ y = 1.0 - a\exp\Bigl[-\Bigl({|x - x_{0}|\over b}\Bigr)^{c}\Bigr] \]

This is the function with the fewest free parameters that gives a good
fit while remaining well behaved. Because the lines have a sharp core,
in fitting we convolve model spectra with a Gaussian of FWHM equal to
the instrument resolution i.e. we measure the deconvolved profile.  The
function has four free parameters. The parameter $x_{0}$ is the
wavelength of the line centre. The parameter $a$ is the depth at the
line centre. The parameter $b$ is the scale length of the exponential,
and so gives a measure of the line width. We refer to $b$ as the {\em
scale width} in this paper, to distinguish it from the other line width
measure we use, $D_{0.15}$. Finally the parameter $c$ quantifies the line
shape. Note that $c=2$ corresponds to a Gaussian, $c=1$ to an
exponential, and $c=1/4$ to the de Vaucouleurs profile (de Vaucouleurs
1948). For the same value of $b$, a smaller value of $c$ produces
broader wings to the line.  An example fit for a typical BHB star in
the KSK sample is shown in Fig. \ref{plot_fits} (b). The shapes of the
Balmer lines are discussed further in \S4.2.4.

We found that the parameter $a$ is not useful for separating the two
populations, but that in a plot of $b$ against $c$ the two populations
are distinct (Fig. \ref{scale_power}). Because of the relatively low resolution of the
KSK spectra (typical FWHM$=4\,$\AA), and the sharpness of the line core,
the (deconvolved) parameter $a$ is difficult to measure
accurately. This is unfortunate because errors on the interesting
parameters $b$ and $c$ are correlated with errors in $a$.  The
parameter $a$ is approximately constant for the strong--lined stars in
the KSK total sample. For this reason we decided to fix $a$ at the
average value $a=0.83$. This is satisfactory for the KSK data, and our
own programme spectra, but might not provide an adequate fit for
higher resolution data with higher S/N. 

To summarize, in measuring the line profiles we compute the
minimum-$\chi^2$ fit of the following function convolved with a Gaussian
of FWHM equal to the spectral resolution:

\[ y = 1.0 - 0.83\exp\Bigl[-\Bigl({|x - x_{0}|\over b}\Bigr)^{c}\Bigr] \]
 
We found satisfactory fits to all the stars in the KSK total sample
with this profile. (This modified function, with $a$ fixed at 0.83, is
not appropriate for B or F stars however.) The {\em Scale
width--Shape} method uses the parameters $b$ and $c$ determined in
this way. It should be noted that the separation boundary would be
slightly different for measurements where $a$ is a free parameter

The measurements were restricted to the $H\gamma$ and $H\delta$ lines.
The $H\epsilon$ line is blended with the Ca II H absorption line, while
the higher--order Balmer lines crowd together and it becomes difficult
to define the continuum. The KSK observations do not cover the $H\beta$
line. The wavelength intervals used in making the fit to the $H\gamma$
and $H\delta$ lines are provided in Table 2. In making the fit
discrepant pixels are iteratively $\sigma-$clipped.  The measured line
width may depend on metallicity due to the presence of weak metal lines
that are not clipped out. In an attempt to reduce the effect of
metallicity we have identified the 10 strongest metal lines that lie
within the profile of either the $H\gamma$ or $H\delta$ line. These are
listed in Table \ref{tab_metal_lines}.  Pixels at the doppler--shifted
wavelength of the metal lines are masked out. We return to the issue of
line blanketing below. The fitting routine runs within IRAF
\footnote{IRAF is distributed by the National Optical Astronomy
Observatories, which are operated by the Association of Universities
for Research in Astronomy, Inc. under cooperative agreement with the
National
Science Foundation.} and is available at {\tt
http://astro.ic.ac.uk/Research/extragal/milkyway.html}.
\begin{table}
 \centering
  \begin{tabular}{@{}lcc}
   \hline \\[-12pt]
	Line  & Restframe wavelength ({\AA}) & Bandpass ({\AA})  \\
	\hline \\[-12pt]
        Ca II K    & 3934 & 3919--3949 \\
	H$\delta$  & 4102 & 4032--4172  \\  
	H$\gamma$  & 4341 & 4271--4411  \\ 
%	H$\beta$   & 4861 & 4741-4981  \\
   \hline
\end{tabular}
\label{tab_bandpass}
\caption{Wavelength regions used in fitting the profiles
of the Balmer lines and measuring the Ca II K EW.}
\end{table}

\begin{table}
 \centering
  \begin{tabular}{@{}lc}
   \hline \\[-12pt]
	Line  & Wavelength ({\AA}) \\
	\hline \\[-12pt]
 Fe I    &  4045.8 \\
 Sr II   &  4077.7 \\
 Fe I    &  4173.0 \\
 Ca I    &  4226.7 \\
 Fe I    &  4271.8 \\
 Ti II   &  4300.1 \\
 Fe II   &  4351.8 \\
 Fe II   &  4385.4 \\
 Fe II   &  4416.8 \\
 Mg II   &  4481.3 \\
  \hline
\end{tabular}
\caption{Restframe wavelengths of metal lines masked out by the
fitting routine.} 
\label{tab_metal_lines}
\end{table}

\subsubsection{$D_{0.15}$--Colour method}
Over the colour range $0<(B-V)_0<0.2$ the Balmer lines are strong and
broad, and A stars can be separated according to surface gravity by
plotting
line width (suitably defined) against colour, as shown for example in
Fig. \ref{colour_width}.

In order to measure the Balmer line--widths most authors have followed
Rodgers, Harding \& Sadler (1981) and used $D_{0.2}$, defined as the
width measured at a flux level of 0.8 of the continuum flux level. KSK
used a similar measure but the line--depth was determined relative to
the bottom of the Balmer line rather than from the level of zero flux.
This definition produces a measure of the width closer to the
continuum, fully exploiting the gravity difference in the wings. The
advantage of this is made clear by the curves plotted in
Fig. \ref{plot_fits} (a). However because the core of the Balmer lines
is so sharp the (convolved) line depth depends on the spectral
resolution. To combine the best features of the two different measures
we have defined $D_{0.15}$ which is the width measured at a flux level
0.85 of the continuum flux level.

In order to measure the parameter $D_{0.15}$ using the Sersic profile
fit, the profile is re--expressed in terms of the three free
parameters $D_{0.15}$, $x_{0}$, and $c$. (For reference, the relation
between the parameters is $D_{0.15}=2b\ln(0.83/0.15)^{1/c}$.) The
$1\sigma$ uncertainty on $D_{0.15}$, marginalising over the other
parameters, was computed by determining the increment $\Delta
D_{0.15}$ that produced $\Delta \chi^2=1.0$ when minimizing on the
other two parameters. We checked the errors by two methods. First we
measured the scatter in repeat measurements of the same star. Second
we took high S/N spectra and created random realizations of lower S/N
spectra, and measured the scatter in the fitted parameters. In both
cases we found good agreement with the errors output by the fitting
routine.

\subsubsection{Scale width--Shape method}
\begin{figure}
\rotatebox{0}{
\centering{
\scalebox{0.40}{
\includegraphics*[20,140][700,700]{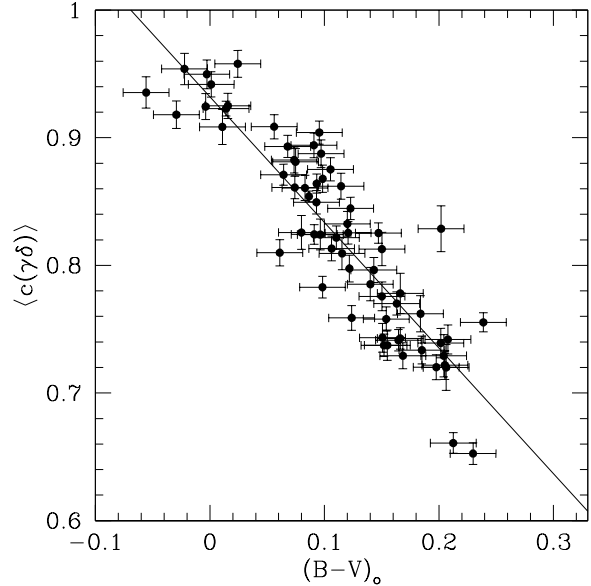}
}}}
\caption{Plot of the shape parameter $c$ (averaged for the H$\delta$
and H$\gamma$ lines) against $(B-V)_0$ for the KSK halo sample,
i.e. A--type stars with Balmer EW$>13\,$\AA. Evidently $c$ provides a
measure of temperature.}
\label{c-colour}
\end{figure}

By plotting different combinations of two of the four parameters $a$,
$b$, $c$, and $(B-V)_0$ for the KSK sample we found that the parameters
$b$ and $c$ provide an alternative method for distinguishing between
BHB stars and blue stragglers. As shown in Fig. \ref{c-colour} over the
colour range $0.0<(B-V)_0<0.2$ the parameter $c$, which quantifies the
shape of the line, is tightly inverse-correlated with colour,
following the linear relation $c=0.93-0.98(B-V)_0$. Evidently $c$
provides a measure of temperature for A--type stars.  At a fixed value
of $c$ (temperature) stars of higher surface gravity have larger
values of the scale width $b$.  In a plot of $b$ against $c$
(e.g. Fig. \ref{scale_power}) BHB stars separate from blue
stragglers.

The differences between the line profiles for the two different types
of star are illustrated in Fig. \ref{plot_fits} (a). Blue stragglers
have larger scale widths than BHB stars of the same temperature.
Alternatively blue stragglers of the same scale width as a BHB star are
cooler (smaller $c$), and therefore have sharper line profile cores and
broader line wings.

The {\em Scale width--Shape} method is effective for stars with strong
Balmer lines i.e.  A--type stars. A major advantage of this method of
discrimination lies in the fact that no photometric measurements are
required to separate the populations.  As $b$ and $c$ are correlated
parameters we compute the error ellipse, marginalizing over the other
parameter, $x0$. In plotting the uncertainties we show semi-major and
semi-minor axes of the ellipse, marking the $68\%$ confidence interval
for each axis in isolation, i.e. marginalizing along the other axis.
We checked the errors in the same way as for the {\em
$D_{0.15}$--Colour} method and again found good agreement.

\subsection{Combining H$\delta$ and H$\gamma$ line
widths}\label{line_syst}
\begin{figure}
\rotatebox{0}{
\centering{
\scalebox{0.40}{
\includegraphics*[20,140][700,700]{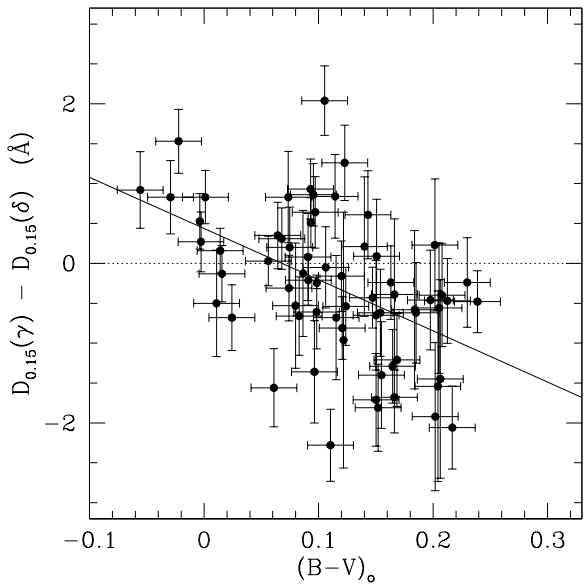}
}}}
\caption{Difference between $D_{0.15}$ for the H$\gamma$ and H$\delta$
lines as a function of colour, for the KSK halo sample. A
systematic difference in line width is evident. A linear
least--squares fit to the data (with 1$\sigma$ errors) gives 
$D_{0.15}(\gamma)-D_{0.15}(\delta)= (0.44 \pm 0.16) - (6.42 \pm 1.17)(B-V)_0$, shown as the
solid line. The dotted line $D_{0.15}(\gamma)-D_{0.15}(\delta)=0.0$ is
also shown as a guide.}
\label{hgam_hdel_col}
\end{figure}

\begin{figure}
\rotatebox{0}{
\centering{
\scalebox{0.4}{
\includegraphics*[20,140][700,700]{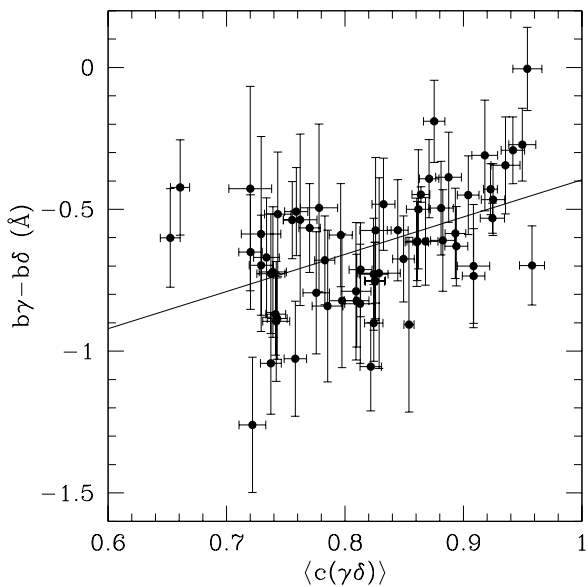}
}}}
\caption{Difference between $b(\gamma)$ and $b(\delta)$ as a function
of $c$, for the KSK halo sample. A systematic difference in line
width is evident. A linear least--squares fit to the data (with
1$\sigma$ errors) gives $b(\gamma) - b(\delta)$ = (1.71 $\pm$ 0.33) - (1.31 $\pm$ 0.40)$\langle c(\gamma\delta)\rangle$}
\label{bgam_bdel_c}
\end{figure}

For each parameter we have measurements for the two lines H$\delta$
and H$\gamma$ in each spectrum. These can be combined to reduce the
errors. For each of the width parameters $D_{0.15}$ and $b$ we found
small but significant differences between the values measured for the
H$\delta$ and the H$\gamma$ lines, with a dependence of the difference
on temperature. This effect is illustrated in Figs \ref{hgam_hdel_col}
and \ref{bgam_bdel_c}. The first of these plots the difference in
$D_{0.15}$ between the two lines against $(B-V)_0$. The second of these
plots the difference in $b$ between the two lines against the
parameter $c$. We have chosen the H$\gamma$ line as our reference, and
we have used the linear fits to calibrate the values of $D_{0.15}$ and
$b$ from H$\delta$ to H$\gamma$ and then calculated weighted
averages. For the parameter $c$ we found no such trend, and so we
simply computed the weighted average of the $c$--values for the two lines.

We have not discovered a convincing explanation for the trends
seen. We suspect it is an effect of line blanketing from weak metal
lines, which would explain the temperature dependence. However if this
is the case one would expect there to be a correlation between the
residual from the fit and the metallicity, as estimated from the Ca II
K line. However we found no significant correlation in a plot of these
two quantities.

\subsection{Ca II K-line EW measurements}
The Ca II K line is the strongest measurable metal line over the
wavelength range covered by the KSK spectra, and the only useful
indicator of metallicity in spectra of low S/N, such as for the stars
in our radial velocity programme. With some exceptions (including the
A metallic (Am) and peculiar (Ap) stars) the strength of the Ca II K
line at constant temperature can be used as a reliable indicator of
the metallicity of A-type stars (e.g. Pier 1983, Beers et al.
1992, KSK). We measure the EW of the Ca II K line by
minimum$-\chi^2$ fitting a Gaussian to the continuum divided spectrum
over the wavelength range given in Table 2. This line is much weaker
than the Balmer lines, and to reduce the error the central wavelength
is fixed at the redshift determined from the Balmer lines.  

In \S6 we use a plot of EW$_{Ca}$ versus $(B-V)_0$ to
estimate the metallicities of the KSK stars.  This has two
purposes. First it allows anomalous high--metallicity halo stars to
be identified. Secondly knowledge of the metallicity can be used to
reduce the uncertainty in the absolute magnitudes of the BHB stars,
and consequently, the errors in the distances (e.g. Clementini et al.
1995). The issue of measuring distances is deferred to Paper II.

\begin{figure*}
\begin{minipage}{175mm}
\epsfig{figure=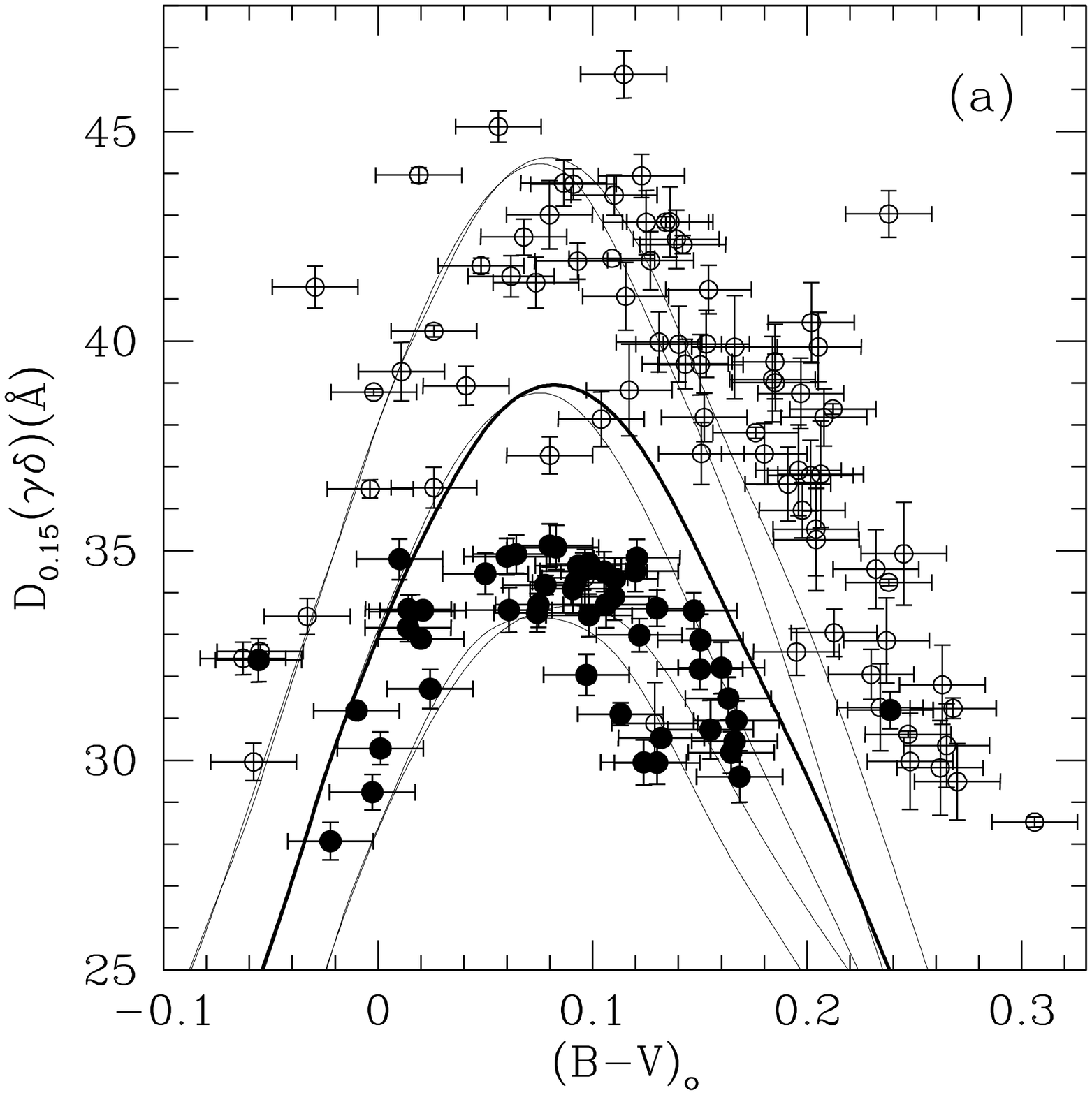,width=86mm}
\epsfig{figure=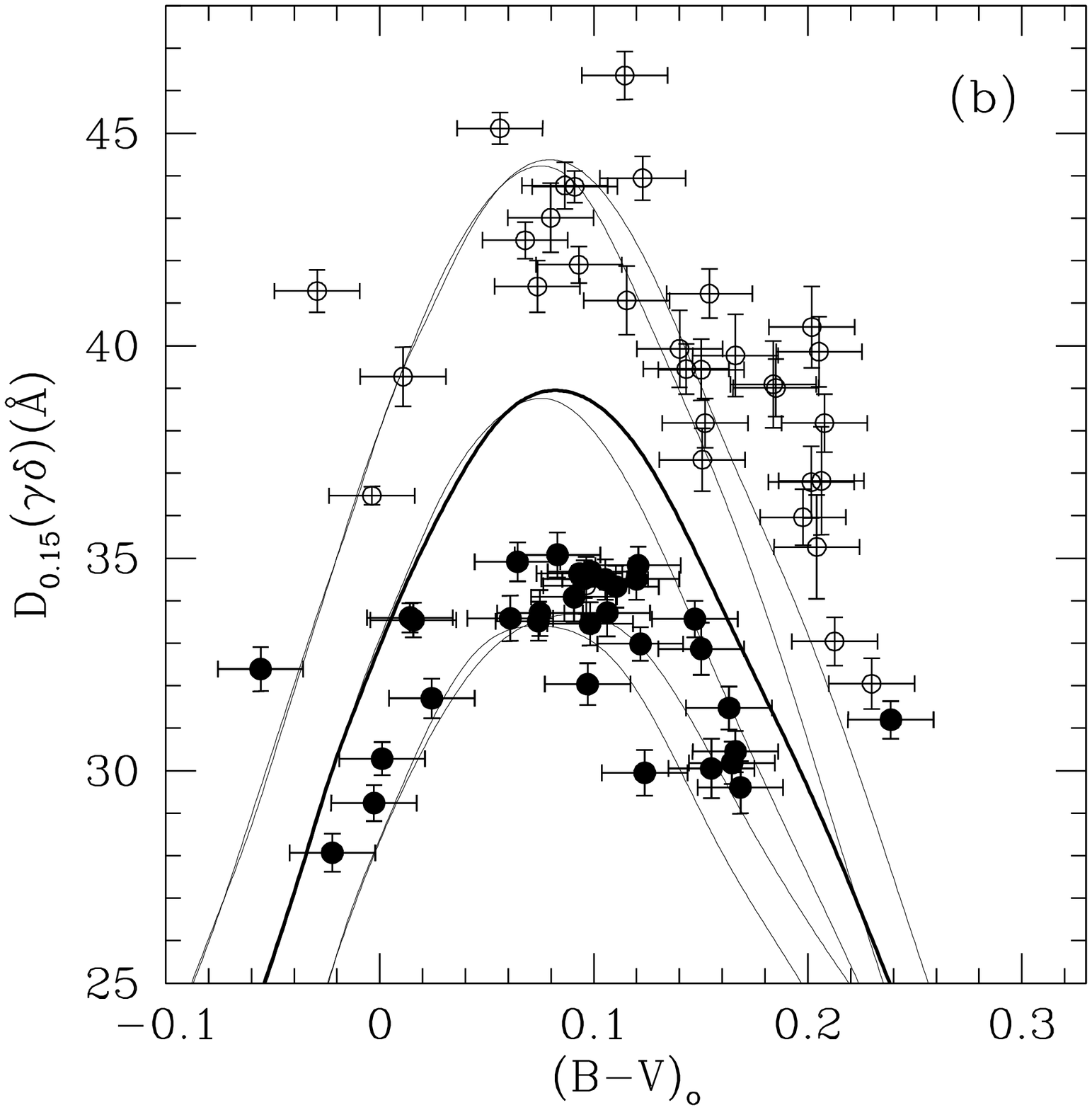,width=86mm}
\end{minipage}
\caption{Separation of the KSK sample using the {\em
$D_{0.15}$--Colour} method: (a) the KSK total sample (131 stars), (b)
the KSK halo sample (66 stars). BHB stars, classified as such by KSK
using their $\Lambda$ method, are marked by filled circles, and stars
classified A/BS are marked by open circles. The curves show results for
model spectra from Kurucz (1993) for log g = 3.0, 3.5, 4.0. Each
gravity has two lines of metallicity [Fe/H] = -2 (left) and -1
(right). The line log g = 3.5, [Fe/H] = -1, plotted bold, is our
chosen classification boundary.}
\label{colour_width}
\end{figure*}

\section{Results: The Balmer lines}

In this section we show how the two classification methods work in
practice. Using measurements of the Balmer lines for the KSK sample we
present plots illustrating the two methods. We then quantify the S/N
required to apply the classification methods to samples of faint halo
A--type stars.

\subsection{{\em $D_{0.15}$--Colour} method} \label{width_colourm}

The {\em $D_{0.15}$--Colour} method is illustrated in
Fig. \ref{colour_width}, which plots $D_{0.15}$, averaged for the two
lines, against $(B-V)_0$ colour. We have used KSK's classification
using the spectrophotometric $\Lambda$ parameter to separate the
samples into two populations. Note that because this method uses the
shape of the continuum it is independent of classification that uses
the line widths. In the two plots stars classified BHB are plotted as
filled symbols and other stars i.e. blue stragglers or main-sequence A
stars, hereafter `A/BS', are shown as open symbols. The left-hand plot
is for the KSK total sample, which includes stars with a range of
metallicities, while the right-hand plot shows the KSK halo sample
only.

\begin{figure}
\rotatebox{0}{
\centering{
\scalebox{0.30}{
\includegraphics*[-50,140][700,700]{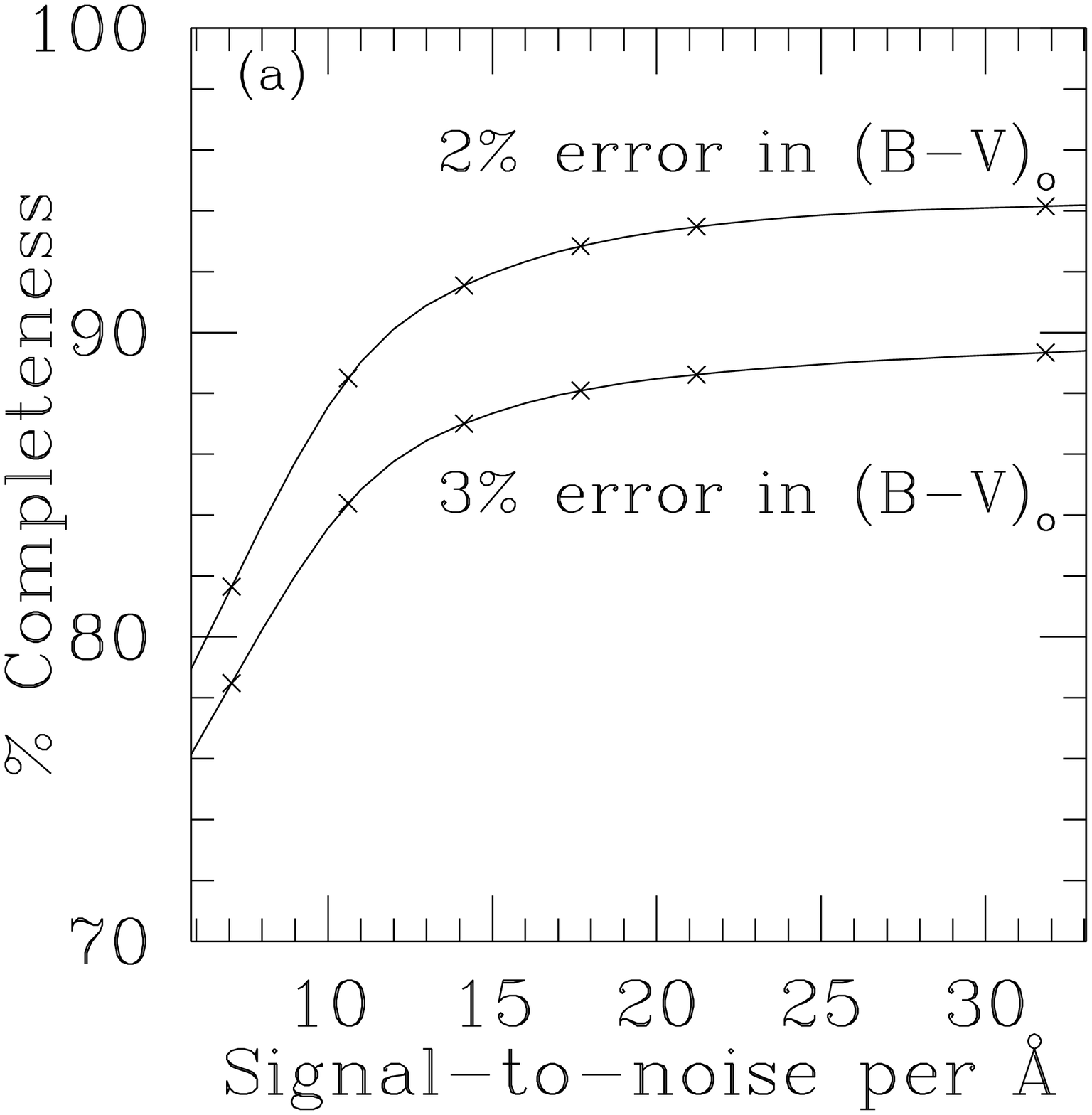}
}}}
%\end{figure}
%
%\begin{figure}
\rotatebox{0}{
\centering{
\scalebox{0.30}{
\includegraphics*[-50,140][700,700]{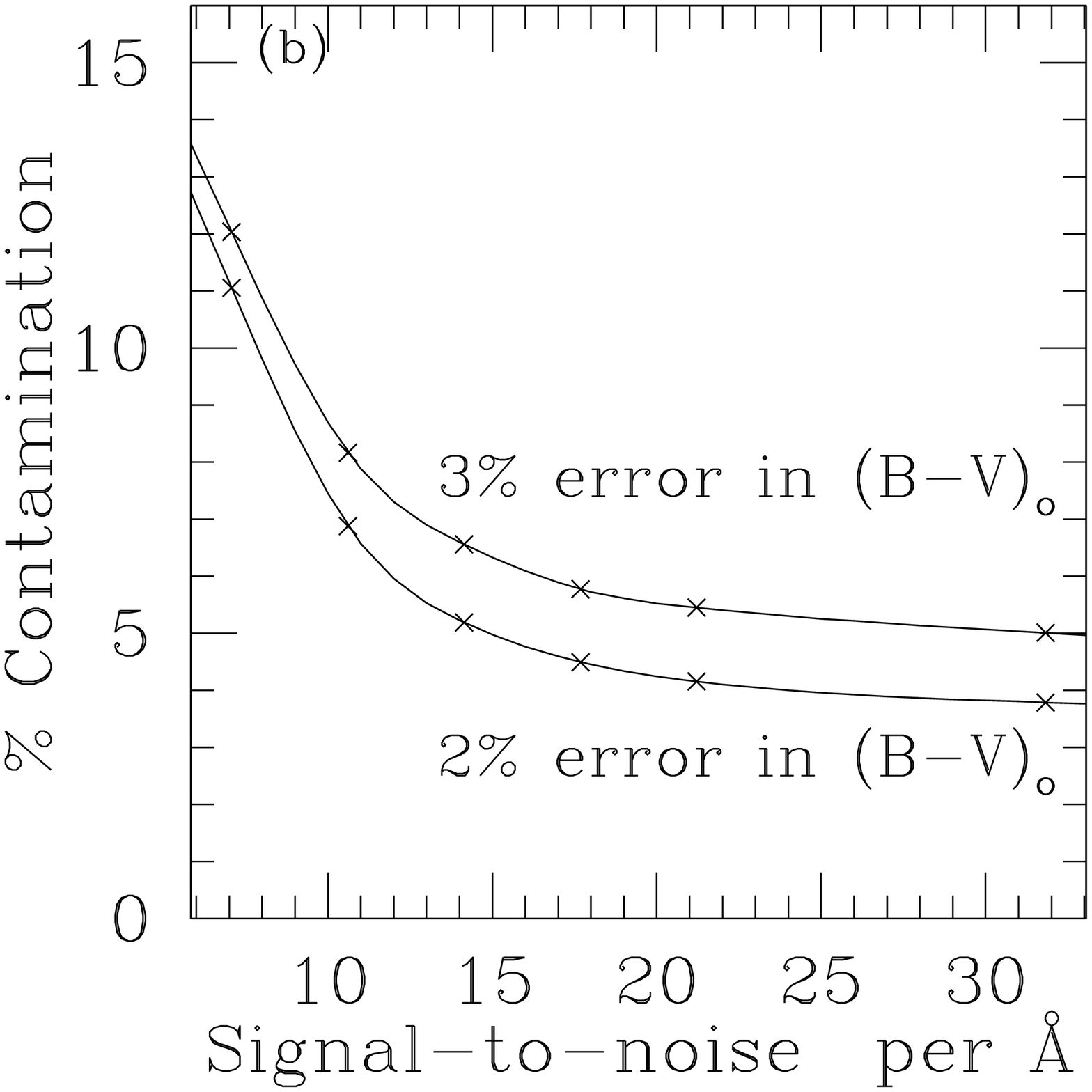}
}}}
\caption{Monte Carlo simulations of the halo sample of stars shown in
Fig. \ref{colour_width}(b): (a) completeness of the BHB sample as a
function of spectroscopic S/N, for 2\% and 3\% errors in the colours
(b) contamination of the sample by blue stragglers.}
\label{col_con_comp}
\end{figure}

The two populations BHB and A/BS are cleanly separated in these plots.
The two sequences stand out more clearly in the left-hand plot
because of the larger number of objects. The curves overplotted are
derived from model spectra. The calculation of the models is described
in the Appendix. Curves for three surface gravities $\rm{log \,
g}=3.0, 3.5, 4.0$ are shown, the line width increasing with surface
gravity. For each value of $\rm{log \, g}$ curves for two different
metallicities [Fe/H] = -2.0, -1.0 are plotted. The right-hand
curve of each pair corresponds to the higher metallicity. While the
curves follow the broad trends in the data, in detail they provide a
rather poor fit. Main sequence stars in this temperature range have
$\rm{log \, g}\sim 4.5$ whereas the curves with $\rm{log \, g}=4.0$
provide the best fit to A/BS sequence near $(B-V)_0=0.1$, while
falling below the data at redder colours.

Since we are interested in classifying samples of faint halo A--type
stars we concentrate on the right-hand plot in
Fig. \ref{colour_width}. We have chosen the curve computed for
$\rm{log \, g}=3.5$ and [Fe/H] = -1.0, marked bold, as the
classification boundary. Above the line there are 31 stars classified
A/BS by KSK and 2 stars classified BHB. Below the line there are 32
stars classified BHB and only 1 star classified A/BS. For reference, in
the left hand plot above the line there are 77 stars classified A/BS by
KSK and 3 stars classified BHB, and below the line there are 47
stars classified BHB and 4 stars classified A/BS. Clearly with data
of this quality highly reliable classification is possible with the
{\em $D_{0.15}$--Colour} method. The nature of the few stars below the
line classified A/BS is discussed in \S6. Most appear to be anomalous by
virtue of their high metallicity and can be identified and removed.

\subsubsection{S/N requirements}
We have used Monte-Carlo methods to assess the spectroscopic and
photometric S/N requirements for classifying samples of more distant
halo A--type stars using the {\em $D_{0.15}$--Colour} method. We take
a figure of $\sim10\%$ as the practical acceptable limit of
contamination of any sample of halo BHB stars by blue stragglers. We
started with the simplifying assumption that the KSK halo sample is
representative of the halo population at fainter apparent magnitudes
in terms of the ratio of blue stragglers to BHB stars. Then, adopting
specific values of spectroscopic and photometric S/N, for each star in
the KSK halo sample we drew a value of $D_{0.15}$ and $(B-V)_0$ from
the error distributions to create a synthetic halo sample. By creating
thousands of samples and counting the number of stars originally
classified A/BS or BHB that appear below the classification boundary in
each sample we have estimated the contamination and completeness of
samples of faint halo BHB stars as a function of S/N.

Fig. \ref{col_con_comp} shows the results of these simulations. The
upper plot shows the percentage completeness of BHB samples as a
function of spectroscopic S/N, for two fixed values of photometric
S/N. The lower plot shows the percentage contamination of BHB
samples. For our faint programme stars $16<B<20$ we have found it
difficult to achieve errors in $(B-V)_0$ smaller than $3\%$. For this
level of photometric accuracy, Fig. \ref{col_con_comp} shows that a
spectroscopic S/N of $15\,$\AA$^{-1}$ would achieve $\sim87\%$
completeness with contamination by blue stragglers at a level of
$\sim7\%$. We consider this acceptable.

Coincidentally this S/N requirement matches the requirement for
measurement of the radial velocity. In our experience with spectra of
S/N=15\AA$^{-1}$ the accuracy of the radial velocity measurement is 15
km s$^{-1}$, which is satisfactory considering that the
one-dimensional velocity dispersion in the halo is $\sim 100$ km
s$^{-1}$.

To summarize, with $(B-V)_0$ colours accurate to 3$\%$, with spectra of
S/N=$15\,$\AA$^{-1}$, samples of halo BHB stars that are $\sim87\%$ complete,
and with only $\sim7\%$ contamination by blue stragglers, can be compiled
using the {\em $D_{0.15}$--Colour} method.

It would be possible to reduce the contamination at the expense of
sample size, by restricting the sample to a narrower range of
$(B-V)_0$ where the discrimination is cleaner. For example if we now
limit the analysis of the KSK halo sample to stars in the range
$0.05<(B-V)_0<0.15$ the number of BHB stars is reduced from 32 to
19. For this sample, with $(B-V)_0$ colours accurate to 3$\%$ and with
spectra of S/N=$15\,$\AA$^{-1}$, compared to the sample with the
broader colour range the completeness is increased from $\sim87\%$ to
$93\%$ and the contamination is reduced from $7\%$ to $5\%$.

\subsection{{\em Scale width--Shape} method} 
\begin{figure*}
\begin{minipage}{175mm}
\epsfig{figure=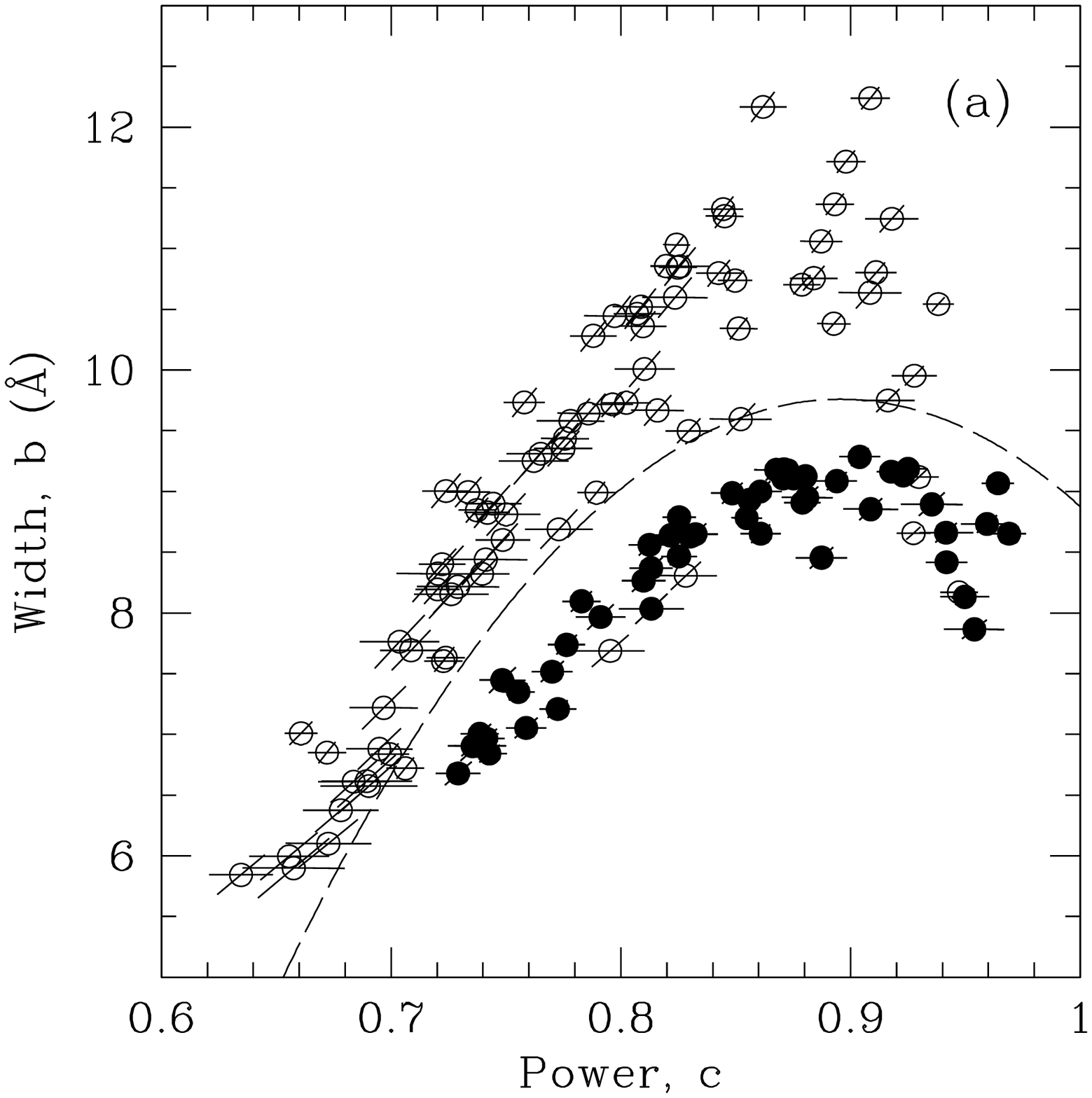,width=86mm}
\epsfig{figure=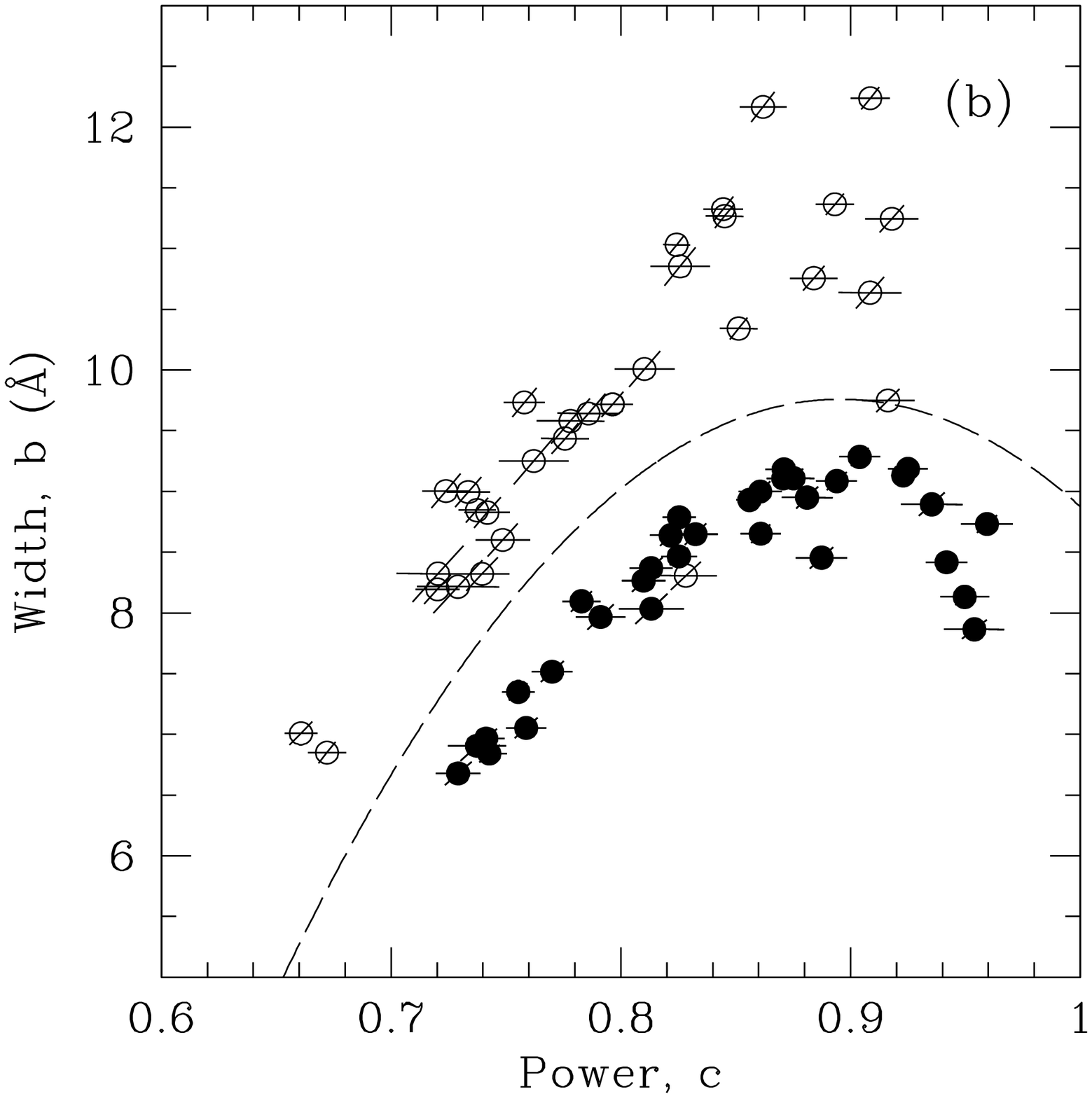,width=86mm}
\end{minipage}
\caption{Separation of the KSK sample using the {\em Scale
width--Shape} method: (a) the KSK total sample (131 stars), (b) the
KSK halo sample (66 stars). BHB stars, classified as such by KSK using
their $\Lambda$ method, are marked by filled circles, and stars classified
A/BS are marked by open circles. Curves showing results for
model spectra provide a very poor fit and have been omitted. The
dashed curve indicates the adopted classification boundary.}
\label{scale_power}
\end{figure*}

The {\em Scale
width--Shape} method is illustrated in Fig. \ref{scale_power}, which
plots $b$ against $c$, averaged for the two lines. As before, the
left-hand plot is for the KSK total sample, while the right-hand plot
shows the KSK halo sample only. Again the two populations BHB and A/BS
are clearly separated in these plots. The curve shown is the empirical
selection boundary that we have chosen for this method. Unfortunately
the model curves are a very poor match to the data and therefore have
not been plotted. The problem appears to be with the shapes of the
lines more than with than the widths. We find that the model spectra
reproduce the slope of the relation between $c$ and $(B-V)_0$
reasonably well, but that the locus for the model spectra is offset to
higher values of the parameter $c$. In other words the Balmer lines of
the models are not spikey enough.

In the right-hand plot, showing  the halo sample, above the line there
are 31  stars classified A/BS by  KSK and 0 classified  BHB. Below the
line  there are 34  stars classified  BHB and  only 1  star classified
A/BS. For reference, in the left hand plot above the line there are 73
stars classified A/BS by KSK and 0 stars classified BHB, and below the
line  there  are  51  stars  classified BHB  and  7  stars  classified
A/BS.  Therefore the  {\em Scale  width--Shape} method  appears  to be
similarly effective in separating  the two populations.  The nature of
the  few  stars  below  the  line  classified  A/BS  is  discussed  in
\S6.  Again, most  appear  to be  anomalous  by virtue  of their  high
metallicity and can be identified and removed.

\subsubsection{S/N requirements}

\begin{figure}
\rotatebox{0}{
\centering{
\scalebox{0.30}{
\includegraphics*[-50,140][700,700]{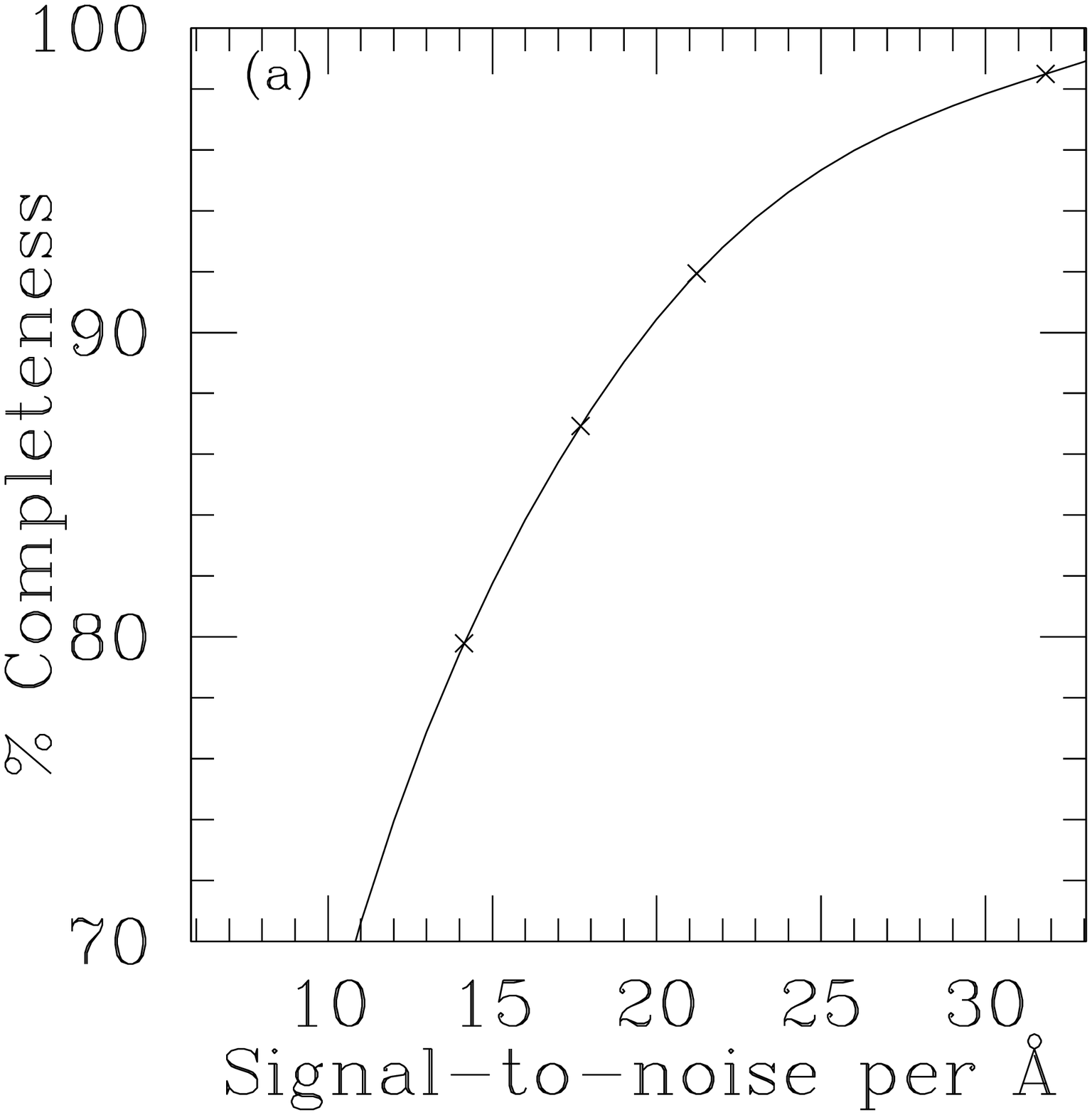}
}}}
%\end{figure}
%
%\begin{figure}
\rotatebox{0}{
\centering{
\scalebox{0.30}{
\includegraphics*[-50,140][700,700]{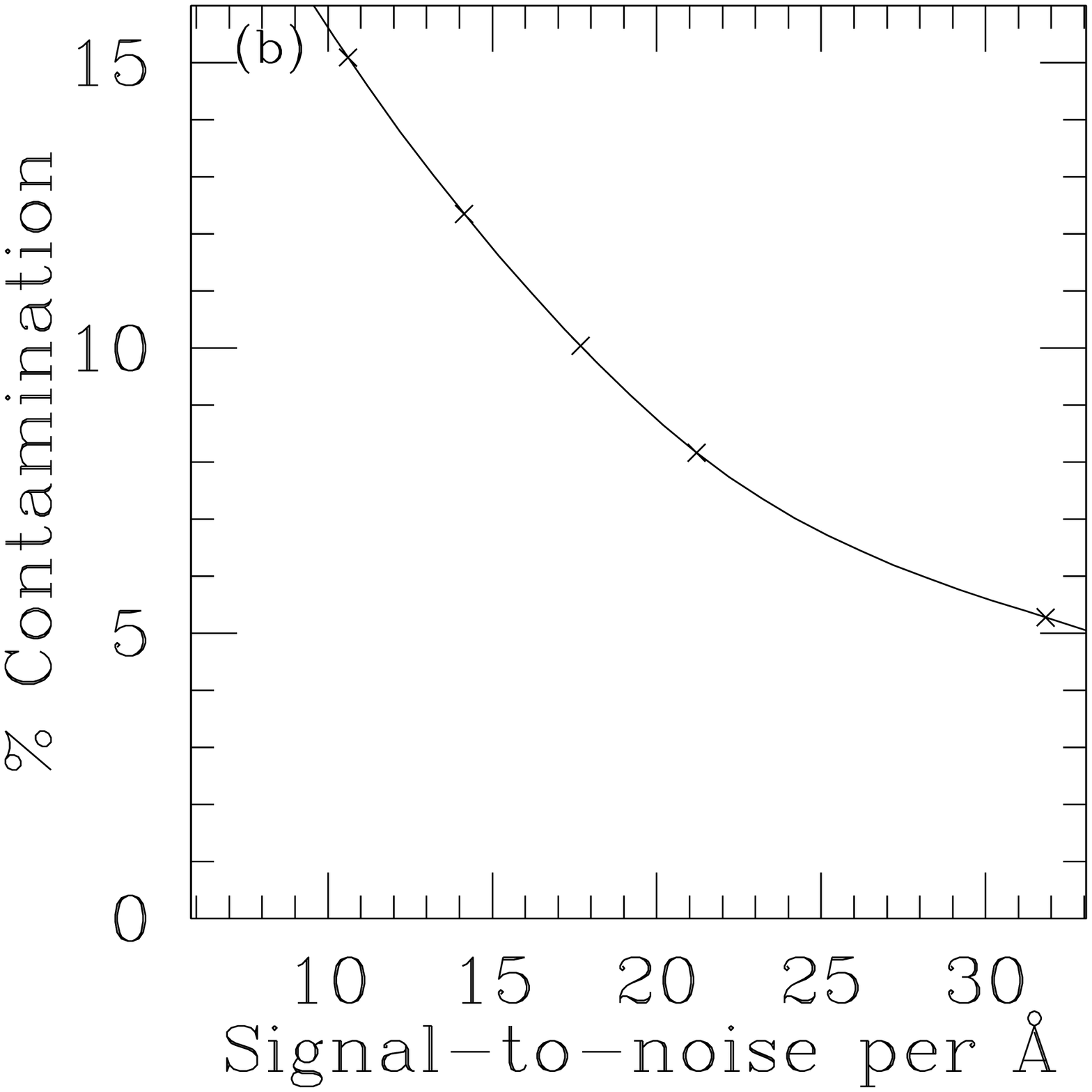}
}}}
\caption{Monte Carlo simulations of the halo sample of stars shown in
Fig. \ref{scale_power}(b): (a) completeness of the BHB sample as a
function of spectroscopic S/N, for 2\% and 3\% errors in the colours
(b) contamination of the sample by blue stragglers.}
\label{bc_cont}
\end{figure}

We have followed the same Monte-Carlo procedure used for the {\em
$D_{0.15}$--Colour} method to establish the S/N requirements for the
{\em Scale width--Shape} method. The results are plotted in
Fig. \ref{bc_cont}. Here the only variable is spectroscopic S/N. The
completeness and contamination for this method are slightly worse than
for the {\em $D_{0.15}$--Colour} method. For a spectroscopic S/N of
$15\,$\AA$^{-1}$ the completeness is $\sim82\%$ (compared to $87\%$)
and the contamination is $\sim12\%$ (compared to $7\%$).

We see that with spectroscopy alone we can separate the two
populations almost as well as with the {\em$D_{0.15}$--Colour} method.
Because accurate photometry is not required there will clearly be
circumstances where the {\em Scale width--Shape} method is
preferred. For example if the photometric accuracy of the parent
catalogue from which the A--type stars are selected is significantly
worse than $3\%$ the colours are not useful for classification. It may
then be most efficient to simply obtain spectra of the candidate BHB
stars. This saves the time required to obtain accurate photometry,
with the only penalties a small reduction in completeness and a small
increase in contamination. This may well be the best strategy for
exploring the outer reaches of the Galaxy's halo using the SDSS
dataset; stars at 100 kpc distance have $g^{\prime}\sim21$, where the
error on the Sloan $g^{\prime}-r^{\prime}$ colour is $5\%$ (Stoughton
et al., 2002).

\section{Results: The Ca II K line}

In this section we establish the accuracy with which metallicities can
be measured using the Ca II K line EW. We then show that the
estimated metallicities are useful for identifying interlopers in the
BHB samples.

\subsection{Measurement of metallicity}

\begin{figure}
\rotatebox{0}{
\centering{
\scalebox{0.40}{
\includegraphics*[20,140][700,700]{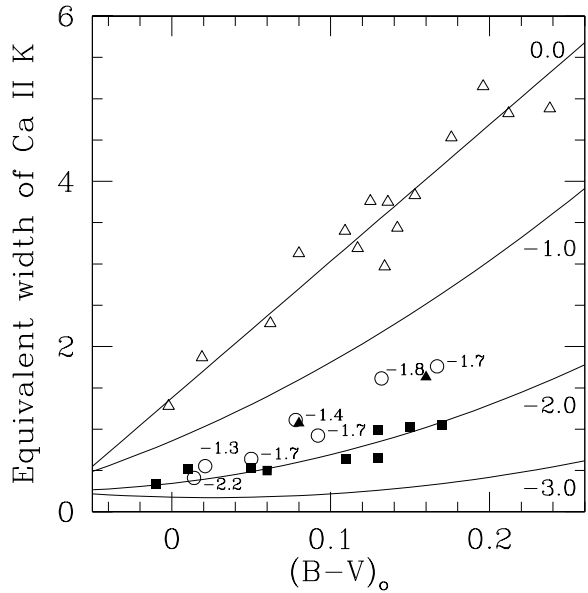}
}}}
\caption{Ca II K line ($3933\,$\AA) EW for selected stars in KSK,
listed in Table 4. The curves represent lines of metallicity for
[Fe/H] = -1.0, -2.0 and -3.0 taken from Wilhelm et al. (1999a). The
straight line represents a best fit to stars in the Pleiades and Coma
clusters (open triangles) assumed to be of solar metallicity. Filled
squares are stars from M92 ([Fe/H] = -2.2) and filled triangles are
from M3 ([Fe/H] = -1.5). Field BHB stars are shown by open circles
labeled with the KSK adopted values.}
\label{cak_plot}
\end{figure}

KSK plotted the Ca II K line EW against $(B-V)_0$ for stars for
which independent accurate metallicities are known, to determine
empirically curves of constant metallicity in this parameter space.
Instead we have used the theoretical curves of Wilhelm et al. (1999a)
measured from synthetic spectra. In Fig. \ref{cak_plot} we plot their
isoabundance contours for [Fe/H] = --1, --2 and --3 (the
solar-metallicity line is explained below). Metallicities can be
determined from this plot using measurements of EW$_{Ca}$ and $(B-V)_0$,
by interpolation. To assess the accuracy of this plot we have used it
to measure metallicities for all the BHB stars in the KSK total sample
for which independent accurate metallicities are known. The stars
used, listed in Table 4 and plotted in Fig. \ref{cak_plot}, are the
BHB stars in M92 ([Fe/H] = --2.2), in M3 ([Fe/H] = --1.5), and a
sample of nearby field BHB stars (individually labeled with the value
of [Fe/H]). The accurate metallicities are taken from KSK and are
listed in Table 4 in the column headed [Fe/H]$_K$.

Our interpolated measured metallicities are listed in the last column
of Table 4, labeled [Fe/H]$_C$. There is quite good
agreement between our interpolated estimates and the accurate
determinations. For the sample of 18 stars we measure a mean
difference [Fe/H]$_C$--[Fe/H]$_K=0.10$ with standard deviation
$\pm0.24$. For the three sub-samples M92, M3, and the field BHB stars
the results are $0.21\pm0.20$ (8 stars), $-0.10\pm0.17$ (3 stars),
$0.06\pm0.27$ (7 stars) respectively. The scatter of the points could
arise from a variety of sources including the difficulty in measuring
the continuum, contamination of the line by interstellar absorption,
and uncertainties in the accurate estimates themselves (see KSK for a
discussion). Therefore in using this plot to measure metallicities of
other stars we add an error of 0.3 dex in quadrature to the random
error, based on the measured scatter.  Additional scatter may be
introduced by the spectral peculiarities of the Am and Ap stars
(discussed in KSK, Wilhelm et al. 1999a, 1999b).

To measure the metallicities of other A--type stars over the full
range of metallicities we have defined also a solar--metallicity
contour by fitting a straight line to the data for main--sequence A
stars in the Pleiades and Coma clusters, plotted as open
triangles. Because the lines are converging towards bluer colours we
consider this plot reliable for estimating metallicities only for
colours redder than $(B-V)_0>0.05$.

\subsection{Identification of interlopers}

We return now to the issue of interlopers i.e. the few halo stars that
are classified by our methods as BHB but classified by KSK as A/BS. We
have discovered that most of these interlopers have anomalously high
metallicity. 

If we consider first the {\em $D_{0.15}$--Colour} method, then for the
KSK halo sample, shown in Fig. \ref{colour_width} (RHS), there is one
interloper amongst the 33 stars below the classification boundary, the
star numbered RR7-70 by KSK. This star is the only star below the
line with estimated metallicity [Fe/H]$>-0.5$. In the KSK total
sample, Fig. \ref{colour_width} (LHS), there are 4 interlopers amongst the 51 stars
below the classification boundary, and similarly they are the only
stars with metallicity [Fe/H]$>-0.5$.

Turning to the {\em Scale width--Shape} method, in the halo
sample, Fig. \ref{scale_power} (RHS), there is one interloper amongst the 33 stars
below the classification boundary, the star RR7-70, the same
interloper in Fig. \ref{colour_width} (RHS). This star is the only star below the line
with estimated metallicity [Fe/H]$>-0.5$. In the total sample, there
are 7 interlopers amongst the 58 stars below the classification
boundary. For three of these we are unable to estimate metallicities
reliably as they have colours $(B-V)_0<0.05$. Three of the remaining
four are the only stars with metallicity [Fe/H]$>-0.5$.

We conclude that there is a very small proportion of stars classified BHB
by our method that are probably not BHB, and that these can be
identified by their anomalously strong Ca II K line at the given
temperature. Applying this additional filter with high S/N data,
samples of BHB stars that are almost perfectly clean and complete can
be produced using the two classification methods discussed in this paper.

\section{Summary}

The purpose of this paper has been to set out efficient methods for
classifying reliably halo A-type stars into the classes BHB and blue
straggler, and to quantify the S/N requirements. We disregarded the
spectrophotometric $\Lambda$ method and Stromgren $u$ photometry as
possible methods of classification because they require too much
telescope time, bearing in mind that additional spectroscopy is
required to measure radial velocities. Instead we have concentrated on
measurements of the Balmer lines H$\gamma$ and H$\delta$ from
intermediate-resolution ($\sim 3$\AA) spectroscopy (not
spectrophotometry). The {\em $D_{0.15}$--Colour} method plots the
width of the Balmer lines against $(B-V)_0$ colour. The main
improvements we have introduced over previous applications of this
method are profile fitting of the Sersic function, and quantification
of the errors. We find that with spectra of S/N$=15\,$\AA$^{-1}$ and
colours accurate to $3\%$ samples of halo BHB stars will be
$\sim87\%$ complete with only $\sim7\%$ contamination by blue
stragglers. Spectra of this S/N provide radial velocities accurate to
15 km s$^{-1}$. The small contamination of the sample can be further
reduced by identifying stars with anomalously strong Ca II K
absorption for their temperature.

\begin{table}
 \centering
 \begin{tabular}{@{}lccccc}
 \hline \\[-11pt]
	Name & $EW_{Ca}$ & $\sigma_{Ca}$ & $(B-V)_0$ &
 [Fe/H]$_K$ & [Fe/H]$_C$\\
	\hline \\[-11pt]
M92 II-23 & 0.50 & 0.07 & 0.06 & -2.2 & -2.1 	\\
M92 IV-27 & 1.03 & 0.07 & 0.15 & -2.2 & -1.9 	\\
M92 S-20 & 0.65 & 0.07 & 0.13 & -2.2 & -2.3 	\\
M92 S-24 & 0.52 & 0.05 & 0.01 & -2.2 & -1.7 	\\
M92 VI-10 & 0.64 & 0.07 & 0.11 & -2.2 & -2.2 	\\
M92 XII-01 & 1.05 & 0.05 & 0.17 & -2.2 & -2.0 	\\
M92 XII-09 & 0.99 & 0.07 & 0.13 & -2.2 & -1.8 	\\
M92 XII-10 & 0.53 & 0.06 & 0.05 & -2.2 & -1.9 	\\
M3 182 & 1.07 & 0.07 & 0.08 & -1.5 & -1.5 	\\
M3 II-11 & 4.05 & 0.13 & 0.42 & -1.5 & -1.8 \\
M3 VI-18 & 1.63 & 0.06 & 0.16 & -1.5 & -1.5 \\
HD 2857 & 1.76 & 0.06 & 0.17 & -1.7 & -1.5 	\\
HD 14829 & 0.41 & 0.05 & 0.01 & -2.2 & -1.9 	\\
HD 60778 & 1.11 & 0.02 & 0.08 & -1.4 & -1.4 	\\
HD 74721 & 0.55 & 0.01 & 0.02 & -1.3 & -1.7 	\\
HD 86986 & 0.92 & 0.02 & 0.09 & -1.7 & -1.7 	\\
HD 109995 & 0.64 & 0.01 & 0.05 & -1.7 & -1.8 	\\
HD 161817 & 1.61 & 0.02 & 0.13 & -1.8 & -1.4 	\\
\hline 
\label{tab_cak}
\end{tabular}
\caption{Ca II K line EW and the calculated
[Fe/H] ([Fe/H]$_C$) compared to those in KSK ([Fe/H]$_{K}$) for stars shown in
Fig. \ref{cak_plot}.}
\end{table}

A second classification technique, the {\em Scale width--Shape}
method, compares the shapes of the Balmer lines by plotting two of the
parameters of the profile fit. We find that with spectra of
S/N$=15\,$\AA$^{-1}$ samples of halo BHB stars will be $\sim83\%$
complete with only $\sim12\%$ contamination by blue stragglers. In
other words without the need for colours the samples are almost as
clean and complete as with the {\em $D_{0.15}$--Colour} method. This
argues that if accurate colours $<3\%$ are available for the sample
the best method is the {\em $D_{0.15}$--Colour} method. However where
the accuracy of the photometry is worse the most efficient method for
studying the dynamics of the Galactic halo would be the {\em
Scale width--Shape} method, saving the need to obtain accurate
photometry. This may be the best strategy for studying the outer halo
of the Galaxy beyond $r=100$ kpc, with candidate BHB stars selected
from the SDSS database.

\section*{Acknowledgments} We are very grateful to Drs Kinman,
Suntzeff, and Kraft for supplying us with their spectra. The authors
acknowledge the data and analysis facilities provided by the Starlink
Project which is run by CCLRC on behalf of PPARC.

\appendix

\section{Model spectra}

We assembled a grid of spectral line profiles using LTE model
atmospheres created by Kurucz (1993) and the spectrum synthesis code
SYNSPEC (Hubeny et al. 1994). We created spectra over the wavelength
range $3500-4500\,$\AA.  Detailed Stark broadening tables were used
for the hydrogen lines and the flux distribution was calculated every
$0.01\,$\AA. We examined models with $\rm{log \, g}=3.0, 3.5, 4.0,
4.5$, and [Fe/H] = -2.0, -1.0, 0.0 in the temperature
range $6750 \le \rm{T}_{eff} \le 12000$ measured in steps of $250\,$K
to $\rm{T}_{eff}= 10\,000$ and $500\,$K thereafter.  The parameter
space corresponds to $-0.1 < (B-V)_0 < 0.3$, appropriate for our
observations. The resultant spectra were measured in the same way as
the KSK data.

\end{document}